\documentclass{sig-alternate}

\usepackage{balance}
\usepackage{comment}
\usepackage{subfigure}
\usepackage{graphicx}
\usepackage{epsfig}
\usepackage{url}
\usepackage{amsfonts}
\usepackage{amssymb}
\usepackage{mdwmath}
\usepackage{mdwtab}
\usepackage{cite}
\usepackage{booktabs}
\usepackage{tabularx}
\usepackage{multirow} 
\usepackage[]{caption}
\usepackage{amsmath}
\usepackage{graphicx}
\usepackage{hyperref}
\usepackage{xcolor}
\usepackage{relsize}

\makeatletter
\def\@copyrightspace{\relax}
\makeatother

\begin{document}

\numberofauthors{1}
\author{
\alignauthor Prasant Misra$^\ddagger$, Yogesh Simmhan$^{\dagger}$, Jay Warrior$^\ddagger$\\
\small(contributors in alphabetical order)\\
\affaddr{$^{\ddagger}$Robert Bosch Centre for Cyber Physical Systems, Indian Institute of Science, Bangalore, India}\\
\affaddr{$^{\dagger}$Supercomputer Education and Research Centre, Indian Institute of Science, Bangalore, India}\\
\email{\{prasant.misra, jay.warrior\}@rbccps.org$\ddagger$ simmhan@serc.iisc.in$^{\dagger}$}
}

\title{Towards a Practical Architecture for the Next Generation Internet of Things}

\maketitle

\begin{abstract}
The Internet of Things, or the IoT is a vision for a ubiquitous society wherein people and ``Things'' are connected in an immersively networked computing environment, with the connected ``Things'' providing utility to people/enterprises and their digital shadows, through intelligent social and commercial services.  
Translating this idea to a conceivable reality is a work in progress for more than a decade.  
Current IoT architectures are predicated on optimistic assumptions on the evolution and deployment of IoT technologies. 
We believe many of these assumptions will not be met, consequently impeding the practical and sustainable deployment of IoT.  
In this article, we explore use-cases across different applications domains that can potentially benefit from an IoT infrastructure, and analyze them in the context of an alternative world-view that is more grounded in reality. 
Despite this more conservative approach, we argue that adopting certain design paradigms when architecting an IoT ecosystem can achieve much of the promised benefits in a practical and sustainable manner.
\end{abstract}

\section{Introduction} \label{sec:introduction}

\begin{figure*}[t]
\begin{center}
\subfigure[Edward Hall's proxemic zones mapped to ``Things'']{\label{fig:}\includegraphics[width=3.3in]{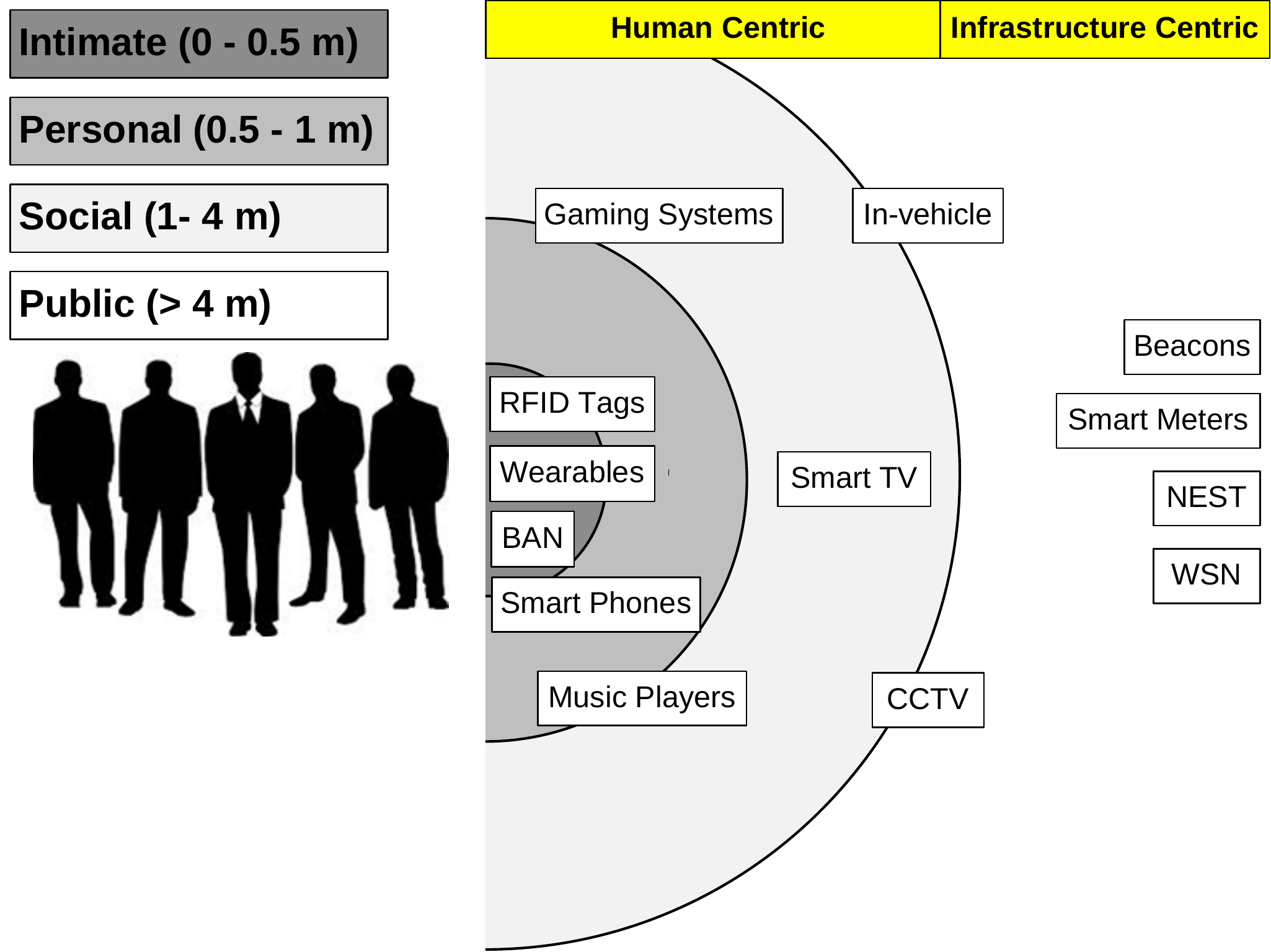}} \hspace{0.5cm}
\subfigure[Coupling vs. Connectivity of ``Things'']{\label{fig:}\includegraphics[width=3.3in]{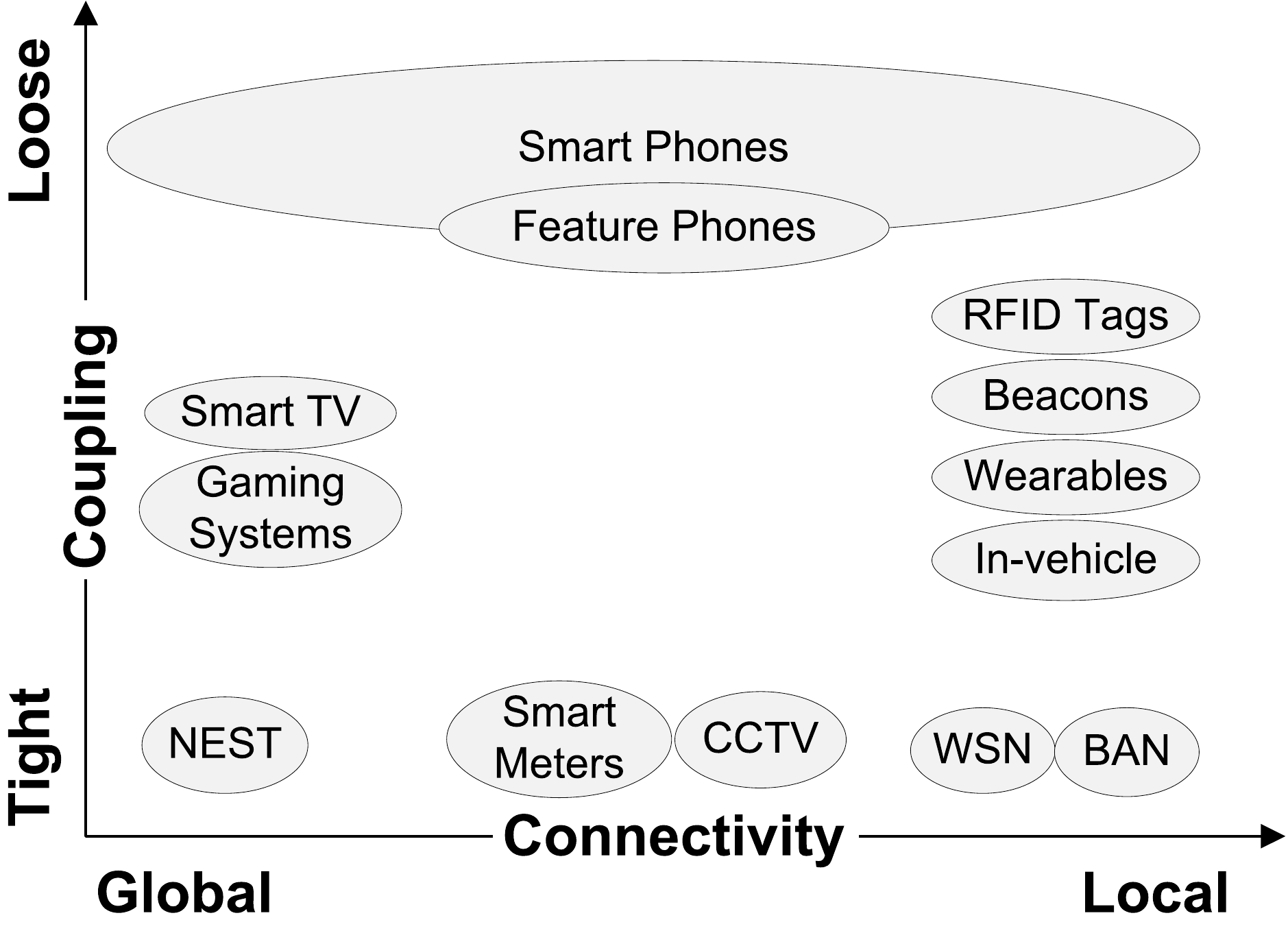}}
\end{center}
\vspace{-5mm}
\caption{People and ``Things'' in the IoT Canvas.}
\label{fig:device-vs-human-proximity}
\end{figure*}

The past couple of years have seen a heightened interest in the Internet of Things (IoT),
transcending industry, academia and government~\cite{IoT-Interest}.  The optimism of IoT being a
\emph{transformational} rather than an \emph{evolutionary} technological advancement is driving
innovation in both traditional and non-traditional application
domains~\cite{traditional-non-domains}.  At the same time, as with promising concepts that are still
emerging, the term IoT is being co-opted by various enterprises and industries to push their own
vision, sometimes molded to fit their existing products and technologies.  IoT is referred to as the \emph{Internet of Everything (IoE)} by Cisco~\cite{cisco-IoT}, the \emph{Industrial Internet} by GE~\cite{GE-IoT}, the \emph{Smarter Planet} by IBM~\cite{IBM-IoT} and \emph{Internet of Things and Humans (IoTH)} by O'Reilly~\cite{oreilly-IoT}.
Notwithstanding these alternate terminologies, the underlying vision -- to build a ubiquitous
society where everyone (``people'') and everything (``systems, machines, equipment and devices'') is
immersively connected -- is over $15$ years in the making \cite{IoT-History}.

As with new ideas that hold immense potential, the enthusiasm exaggerates and outruns the underlying
technology on the ground well before it matures into a sustainable ecosystem. Estimates on the
spread and economic impact of IoT over the next few years are in the neighborhood of $50$ billion or
more connected ``Things'' with a market exceeding \$350 billion~\cite{economic-impact}. The enterprise,
societal and individual benefits touted are also significant, with smarter cities and
infrastructure, intelligent appliances, and healthier lifestyles in the
offing~\cite{societal-impact}. While some of these potential benefits from IoT are real and
achievable, the road to accomplish these is fraught. Proposed IoT architectures typically end up
repackaging existing ideas to benefit from the hype, or assume clean-slate and
costly designs~\cite{IoT-hype-reality} that may prove infeasible. These IoT architecture make assumptions such as:
\begin{itemize}
\item Contain hundreds of devices as part of a tightly coupled infrastructure
\item Deploy devices that cost \$5-500, and are customized for a specific IoT application
\item Use well-structured communication networks (IPv6) with always-on connectivity
\item Perform Cloud-centric data collection and analysis, with centralized control
\item Have a single vendor who owns the platform, Cloud services, data and eco-system for an application
\end{itemize}
Examples of these vertically integrated IoT silos include smart power meters, supervisory control and data acquisition (SCADA) systems, personal monitoring devices like FitBit, etc.,.


An effective architecture for IoT (particularly, for emerging nations with limited technology penetration at the national scale) should be based on tangible technology advances in the present, practical application scenarios of social and entrepreneurial value, and ubiquitous capabilities that make the realization of IoT affordable and sustainable. 
A rethink of the above assumptions suggests that a more plausible scenario would include:
\begin{itemize}
\item Thousands of loosely connected devices in immediate vicinity, and millions more further out
\item Devices that cost from \$0.01-\$3 combined with existing in-person generic devices like smart phones
\item A mix of ad hoc P$2$P, $2$G/$3$G/$4$G and WiFi based on existing and emerging communication networks, and with intermittent connectivity
\item Data collection and personalized analytics that seamlessly span edge devices and the Cloud, with control over data sharing and ownership while encouraging Open Data
\item Open ecosystem without vendor lock-in using standard Internet and Web protocols, allowing devices and data to be shared across IoT applications
\end{itemize}

Humans, data, communication and devices play key roles in the IoT ecosystem. Humans are
an integral part of IoT, which differentiates it from traditional industrial automation or
machine-to-machine (M$2$M) infrastructure.  Humans are not only directly benefited by IoT
applications, but are active participants, serving directly or indirectly as sensors and actuators.
IoT is indeed helping us progress to a point wherein each physical ``Thing'' is a digital shadow of
the person using its services, just as the social network era ushered in a person's data footprint
on the Cloud.  Connected ``Things'' provide utility to people, and their digital shadows provide
value to the enterprise.  Unsurprisingly, various enterprises and companies are working towards linking
``Things'' as diverse as smartphones, cars, smart appliances, etc., with each other, and also with
sensors/actuators~and~the~Internet.


As we consider the broad range of platforms such as RFID tags, beacons, sensor nodes, smart watches, activity trackers, smart phones, in-vehicle sensing systems, and smart meters; it helps to classify them along key dimensions.  One way
is using Edward Hall's theory of proxemics\cite{Marquardt2012:IDP} to classify these according to the
``Things'' spatial closeness to humans, as \emph{intimate}, \emph{personal}, \emph{social} and
\emph{public} (Fig.~\ref{fig:device-vs-human-proximity}).
``Things'' that are closer to humans often provide more insight, are ubiquitous
and less tied to specific infrastructure.  At the same time, we can also classify these ``Things''
based on how coupled they are to specific applications or infrastructure deployments, and how connected they are (e.g., global vs. limited, local and intermittent network connectivity).

On the extremes of the coupling/connectivity quadrant are: (i) beacons that
are very loosely coupled but 
have local connectivity, and (ii) smart meters that have global connectivity to their application, but are tightly coupled.  
The smart phone stands out as it is both loosely coupled with global
connectivity, and also is within the personal proximity range of people\cite{Khan2013:mobile-phone-survey, Dey2011:GCE}.  It can not only function
as the edge device, but also as the gateway to make the necessary interconnection between
an edge device and the Internet.

In the rest of this article, we discuss several usecases for IoT 
and analyze them in the above context (\S~\ref{sec:application-use-case}), offer ten design commandments for a practical IoT architecture
(\S~\ref{sec:paradigms}), suggest a high level IoT reference design that captures these concepts
(\S~\ref{sec:next-gen-architecture}). 



\section{Application Domains} \label{sec:application-use-case}

In this section, we will briefly discuss existing IoT applications, which provide a basis for illustrating various challenges in the current IoT architecture.

\subsection{Smart Healthcare}

The healthcare sector is a fast growing IoT market with a host of new tools and technologies\cite{medcitynews}.
It is expected to improve the quality and efficiency of various healthcare services, but at reduced costs.
The early applications of IoT in healthcare were based on tracking systems for: hand-washing, extracting the location footprints of patients and doctors, monitoring medical assets, and the management of medical wastes\cite{Bureau2010:RFI}.
The tracked information is coupled with contextual information to raise alerts and triggers, facilitate interventions, and assist in identification and verification.
\newline
\indent
Existing tracking systems, for healthcare applications, are designed to be \emph{minimally} invasive\cite{Pantelopoulos2010:healthcare-wearables}. 
They were architectured in a manner such that: $1$) a lightweight passive tag is carried by the object-of-interest; $2$) readers/interrogators, typically, fixed to some location in the defined tracking perimeter respond to tag information; and $3$) a central server subsequently analyzes the tagged information to deriving useful analytic. 
Such architectures are examples of special purpose (tracking) systems with vertically integrated operation scope; and hence, introduce certain constraints.
A close look leads to the following indications of a possible alternative architecture.
\newline
\indent
\emph{First}, existing solutions use special purpose tracking hardware and tags. 
While RFID tags may be commoditised, they are still a special device that is out of place in daily life.
Given the prominence of smart gadgets (such as smartwatches) and portable devices (such as smartphones) that are inevitably carried by people and are part of their natural ecosystem, there exists the potential to replace these custom hardware by `symbiotic' usage of existing devices.
\newline
\indent
\emph{Second}, existing IoT architectures default to a centralized model of state space tracing wherein all data collected by the tag reader on an object-of-interest is streamed, and aggregated at a single location for processing and analysis.
In such a case, the scalability of the solution would become a bottleneck as the number of tracked devices increases.
On the contrary, using devices such as smart phones and gadgets as the tracker could potentially resolve this problem; wherein, each such tracker could cater to a group of tracking entities, perform various computations locally on the devices, and those trackers (potentially much smaller in count) could only feed in the required information to the central server.
\newline
\indent
\emph{Third}, the human factor seems to have been omitted in the solution design space; and therefore, they do not tend to take advantage of the available information/features that humans and their motion captures.
In this regard, certain degree of `symbiotic' knowledge could be used wherein we could potentially use people as mules, and use their personal phones/gadgets as the tracker to map out the location beacons instrumented in the space-of-interest; and track their location individually.

\subsection{Smart Energy and Power Grids}

Smart power grids are a canonical cyber physical system. 
Monitoring capability at the utility scale, ranging from phasor measurement units (PMU) at the transmission network to smart power meters at the consumer-end in the distribution network\cite{Kazmi2014:RWB}, are connected through bi-directional networking with the power utility. 
Utilities leverage the energy consumption data, collected through monitoring, to perform decisions on demand-response optimization; and, by using the communication capability, broadcast notifications to consumers (such as request for load shedding/shifting and updates on real-time pricing for the service area). 
In automated demand response systems, these notifications may directly reach the local building control system that enacts energy curtailment routines to meet the utility's request. 
Such smart power grids exemplify the concept of vertically integrated and wide-area distributed IoT; but, the potential for smart energy management goes well beyond the existing mechanisms.
Looking at this architecture differently, we again see indications of a possible alternative architecture.
\newline
\indent
The \emph{first} aspect is to make the energy demand and curtailment process more customer focused rather than utility focused. Understanding the impact of user activities on energy use is important. 
For example, the evening energy consumption can get impacted by a significant delay in the evening traffic on a commuter road.
Likewise, similar energy correlations are expected if consumers are planning to go out for an evening movie. 
Further, when a peak load is detected, utilities need to intelligently target customers for curtailment rather than broadcast to all in the service area, based on the user's past response. 
The incentives given to customers may also vary, appealing to their budgets in one case or environmental consciousness in another. 
Large scale data analytics over historical energy usage to predict demand, mining social network feeds for understanding preferences, and behavioral modeling to determine responses will benefit in these cases.
This requires treating the \textit{IoT as the IoTH}, placing user signals and information streams on par with physical sensing of infrastructure.
\newline
\indent
The \emph{second} aspect is leveraging local rather than centralized control. 
Here, the consumer's software agent acts on their behalf to analyze (rich) personalized information about the user (comfort preferences, activity, schedule, etc.,), building and local weather conditions, along with utility requests to determine the current energy configuration. 
Devices like Nest\cite{googlenest} are early steps in this direction. 
Also, this allows the local agent to \textit{negotiate with the utility} to make trade-offs, thus enabling demand-response optimization as a regular matter rather than as a special scenario.
\newline
\indent
The \emph{third} aspect is one of data ownership. 
One extreme is data siloing, where the utility has only smart meter and smart grid data, and the customer has smart meter and residential/personal data. 
Here, agents for the utility and customer operate on limited data. 
At the other extreme, a utility may collect in-depth data on smart appliances and demographic indicators on the user that while helping with better model prove invasive to privacy. 
Instead, starting with data silos where data owners hold their data but \textit{adding a data exchange and brokering layer} that allows select data to be exchanged with the other party, in return for incentives, would retain control while allowing for optimizations. 
For example, a utility may offer a discount on a customer's bill if they are willing to share plug load information. 
This is also a more nuanced goal than a simple opt-in or opt-out mechanism, with longevity and incentives factored in the end use of the data.

\subsection{Smart Agriculture}

The agriculture sector is the predominant consumer of water worldwide with the bulk of the water demands directed towards irrigation.
Irrigators in many countries are facing a number of challenges such as a growing shortage of available water, increasing energy costs for pumping, and tighter environmental regulations. 
The combination of these factors, along with the inefficient usage of irrigation water, is starting to impact on agricultural productivity, sustainability and food security of many countries, both developed and developing\cite{Jackson2009:water-energy-nexus}.
\newline
\indent
Recognizing the need for improving the efficiency of water usage, many irrigation systems have been proposed.
Irrigators are increasingly encouraged to adopt micro-irrigation techniques with pressurized irrigation systems
over traditional and less efficient methods of flooding and surface irrigation.
Pressurized irrigation systems consist of a network of pipes to carry water under pressure
from the source (e.g., the pumping station) to the destination (i.e., the irrigation area). They offer the advantages
of minimal water utilization (by pressurized flows) for irrigating large areas, prevent water losses (by seepage, evaporation, etc.,), and can be automated. However, on the flip side, system procurement and operation is costly with higher energy costs.
Maintaining such a pressurized water system with limited electrical supplies (as in developing nations), or soaring energy costs (as in many developing/developed economies) poses a serious challenge.
\newline
\indent
Smart agriculture is aimed towards addressing the challenges of synergy among contemporary technologies and intelligent systems to enhance the resource use efficiency with minimal negative trade-offs, or to advance the development strategies. 
Many ICT solutions have been developed to address these problems\cite{Corke2010:env-wsn}. 
For instance, remote sensing of the soil moisture through network of sensors and integrating time irrigation scheduling to match the crop demand have been demonstrated; but except for larger commercial agricultural operations, large scale adoption has not occurred.
The reasons behind the low penetration of ICT technologies in agriculture reveal interesting facts; and perhaps, suggest the need for an alternate architecture.
\newline
\indent
\emph{First}, custom designing of enabler technologies and knowledge building amongst the stakeholders is lacking due to poor participation of the stakeholder farmers.
Part of the reason for the lack of due diligence being that the land-users and farmers are interested to be part of the eco-system and understand how the ICT system interprets the farming conditions and what suggestions (in the form of actuation commands) does it issue to alter the conditions; but, existing solutions keep the farmer out of the loop.
\newline
\indent
\emph{Second}, the cost of building and maintaining customized infrastructure (such as wireless sensor networks) is too high for any farmer to afford the services (even after amelioration on a large scale).
Besides, it is an extremely cumbersome (and practically infeasible) task to instruments farms with customized WSNs over an entire region, province or state.
Hence, there is a need to use/develop alternate solutions that can potentially piggy-back on existing infrastructure to reach the envisioned scale.

\subsection{Smart Retail}

Retail stores have been early users of large scale data warehouses to analyze customer purchasing behavior, to help with inventory control, discounting and product placement in physical stores. 
Loyalty cards and billing have been key sources of such information.  
\newline
\indent
There is emerging recognition of the role of IoT in smart retail. 
In an era where customers can search on their smart phones for the cheapest price for an item and their attention being distracted by multiple screens, smart retail needs to span the physical and digital worlds, and span the shop floor and the residence\cite{retail}. 
Retail business, in that sense, has been customer centric; and IoTH offers the chance to further personalize this practice and evolve with technology. 
\newline
\indent
Retail pricing is becoming more dynamic and personalized, based on factors such as inventory, current lowest physical and web pricing, and whether this is a planned or impulse purchase. The former requires realtime analytics on online trends while the latter requires customer modeling.
\newline
\indent
On the shopping floor,  sensors (such as iBeacons) that detect customer proximity to goods, smart shelves that sense products being taken (or put back), and smart carts that detect items placed in the cart, need to be leveraged. 
Tasks such as automated billing and inventory restocking are obvious. 
In addition, real time recommendations on additional purchases based on current cart contents, proximity alerts from shelves based on shopping list available on smart phone, and discounts or health feedback offered to encourage purchases are possible. 
\newline
\indent
Amazon has transformed e-commerce, making virtual shopping from the home (or from anyplace) a common transaction mode. 
Smart appliances keeping track of household usage and automatically reordering items going low is not far-fetched. 
Smart TVs will make it as easy as a click to order products whose advertisements are displayed. 
Going further, analytics could allow vendors to preemptively ship products to customers before they order it (``anticipatory shipping''\cite{amazon-shipping}).
\newline
\indent
Such real time, easy, preemptive and targeted purchasing does bring up significant concerns on data privacy.
However, more than other application domains, the focus on people is making retail a testing ground for new IoT ideas. 

\subsection{Environmental Monitoring}

Understanding the natural environment is also an active application space for IoT, and related ICT technologies.
This domain of mapping large scale environmental phenomenon is broad, and includes monitoring of air and water quality, micro-climate in urban regions and rain-forests, biodiversity, cattle, wildlife habitat, etc.,\cite{Corke_pieee10}.
\newline
\indent
Several monitoring solutions have been developed over the years; where the primary idea has been to perform sensing and large scale data
collection of the `parameters-of-interest', and subsequently aggregating these data points in a central place to perform analytical studies and run live information feeds to a web portal.
The core technologies used for sensing and data collection have been either: (i) remote sensing with geonavigation satellites; or (ii) deployment of custom designed wireless sensor networks; or (iii) using mobile sensor nodes mounted on moving vehicles (such as buses, trams\cite{opensense}, bicycles\cite{Eisenman2010:bikenet}). 
All of these approaches result in varying levels of temporal/spatial granularity and coverage; and offer different levels of system complexity, manageability, scalability, and cost.
\newline
\indent
Mobile crowdsourcing\cite{Ganti2011:mobile-crowdsourcing}, using existing infrastructure and ubiquitously used platforms along with ``humans'' as the data mule\cite{Campbell2006:PUS, Stevens2014:citizenscience}, for monitoring in urbanized settings has emerged as an alternate architecture for meeting the price and performance points where they can be used on a large scale \cite{Hull2006:Cartel}. 
Thus, it replaces the physical infrastructure of previous solutions with ``symbiotic'' (or existing, interdependent) infrastructure of the locale and its inhabitants to dramatically lower the cost and simplicity of deployment of the monitoring solution.
However, a pivotal challenge in this architecture is to incentivize the citizen to participate in the data gathering in exchange for relevant services\cite{Rula2014:NOF}.
Similar to smart retail, the focus on people is driving the ecosystem of new IoT ideas for environmental monitoring in (sub)urban settings.

\section{{Nexgen I{\smaller o}T:} Design Paradigms} \label{sec:paradigms}

Humans, data, communication and devices play key roles in the IoT ecosystem.
Based on the use cases explored in the previous section, we can synthesize several characteristics that are desired in a practical architecture for the next generation of IoT.

\subsection{Human-centric rather than Thing-centric}

Current IoT architectures are device or network oriented due to their operational significance in an IoT system.
However, two key aspects that are often ignored are the \emph{humans} who are part of this ecosystem and the \emph{context} within which interaction between people and ``Things'' take place. 
\newline
\indent
The human-centric model of IoT, also referred to as Internet of Things and Humans (IoTH), resets the focus from what is enabled by networked devices and sensors to specifically what benefits can be gained \emph{for} humans who are part of, affected by and influence the network.
There will be a tremendous focus on neighborhood and locality by the way people perceive, interpret and use their personal space.
Here, the richness of information will be more profound at close proximity to ``people'', but will deplete and become less interesting and useful as the proximity zone expands.
For such systems, all related technologies and services have to be people centric such as networking, communication, decision making, quality-of-service, etc., for creating a IoT experience that deeply engages with people.
\newline
\indent
Such an approach has several consequences. 
For example, the sensing and actuation of physical infrastructure can naturally extend to people themselves, with human beings acting as data sources (e.g., tweets, health monitoring), and likewise responding to controls (e.g., notifications to voluntarily turn down the air conditioner to save energy).
This offers the possibility of offering uniform control/service and data plane abstractions to both humans (virtual sensors /actuators) and ``Things''.
In addition, the interface to the physical world needs to be simple and easily understandable for managing the scale, multitude and heterogeneity of devices.


\subsection{Span Virtual and Physical Worlds}

Much of the IoT conversation is about the physical infrastructure and its optimization. 
Bringing in humans and social elements (with their virtual online avatars such as social networks and virtual agents) helps span the digital and physical world, and also integrate across humans and infrastructure. 
Capturing proximity and interactions between humans and ``Things'' (H$2$H, H$2$M, M$2$M), both in the physical and virtual worlds, is necessary for \emph{actionable} intelligence. 

\subsection{``Big-Little'' Data}

Analytics performed on information from diverse sources within the IoT architecture helps with data-driven decision making.
There are two classes of such data: ($1$) transient sensor and personal data collected continuously from humans/physical devices, i.e., ``Little'' data; and ($2$) persistent knowledge-bases and archives that span domains and available in central repositories/Clouds, i.e., ``Big'' data. 
Meaningful analytics requires both Big and Little data to be \emph{combined}, and often in real-time.
\newline
\indent
The analytic services themselves could either be deployed on-demand on devices or Clouds, or be restricted to central data centers. 
Consequently, neither a fully federated nor a completely centralized data storage model will scale or sustain.
An asymmetric storage and service model that adapts to the social and service context is necessary. 
The would be many entities in this mix: 
(a) data generators with realtime, cached or archived data (e.g., from sensors) stored at the source site (e.g., mobile phone);
(b) data owners who can release all/some of this data outside the source site, with/without anonymization, in return for payment or incentives;
(c) knowledge-bases, that have (large-sized) open or proprietary corpus of information that are stored at archive sites (e.g., public Clouds);
(d) integration services that consume data from generators and knowledge-based to offer homogenized information on which analytics can be performed. The services themselves may be able to run on a single site (e.g., mobile phone, private Cloud) or across multiple sites (e.g., P$2$P network), and the integrated data itself can be centralized or distributed. 
Integration services will also help discover data and be willing pay for consuming it.
Another entity would be:
(e) analytic services that consume the integrated data and generated operational intelligence. 
The analytics themselves may run centrally (e.g., on a phone/public Cloud) or distributed (across a P$2$P network), and the result they generate is shared with the analytics consumer. 
Analytics services will be willing to pay for the integrated data they consume, and may charge consumers of the analytics.
The final set of entities would be: 
(f) data brokers that help with data discovery and payment of data owners for consumed data, and negotiation of data movement;
(g) service brokers that help with the discovery of analytics or integration services and their payment, and negotiation of service deployment/movement;
(h) analytics consumers who are willing to pay for the analytics service.
At the same time, data privacy and ownership start becoming concerns as the data moves further away from its source (owner).

\subsection{Analytics from the Edge to the Cloud}

Related to Big-Little data is performing distributed analytics and decision making. The current model of pushing all data to a \emph{central} Cloud for analytics will not scale, is inefficient, and raises privacy concerns. Given the enhanced capabilities of edge devices like smart phones coupled with intermittent network connections, decisions on whether a subset of the Big data and decision-analytic should be pushed to the phone, or the Little data and analytic aggregated in the Cloud have to be automated. These are informed by the device capability, privacy needs, energy and network costs, and application QoS.

\subsection{Bring the ``Network to the Sensor''}

As tens of thousands of \emph{cheap} (few cents) IoT devices proliferate, they will be even more constrained in energy and communication capabilities.
%
Rather than rely on massive deployment of custom sensor networks and new standards, there is value in \emph{piggybacking} on existing, widely adopted standards and reusing symbiotic infrastructure to achieve the system scale and densities at affordable costs.
For e.g., using smart phones as P$2$P data mules for last mile connectivity to sensors, combined with highly functional gateways and Clouds for coordination (rather the centralized storage and control), suggests an \emph{asymmetric} architecture.
\newline
\indent
Given the large diversity of ``Things'', it will be unlikely that they all will share a single type of communication infrastructure. 
It is here that the existing Internet Protocol (IP) will provide the necessary interoperability for glueing existing networks that run on different types of communication links.
IPv$4$ has been the ubiquitous stack for the Internet, and therefore, provides a ruggedly tested framework for versatility, scalability and configuration management.
IPv$4$ will undoubtedly remain in operation for years to come; but given the large scale proliferation of ``Things'', its capability to handle the address space could get challenged.
Therefore, an early adoption of the next version of IP (i.e., IPv$6$) may be a plausible direction.
Due to the highly asymmetric architecture, it may be an overkill to provision for IP in the cheap sensing platforms; but would be a good move to limit it to the highly functional gateways.

\subsection{How ``Low'' can you go ?}

Technology penetration has not been uniform across countries, regions, or, for that matter, industries.
This disparity is a reflection of the differences in infrastructure, cost of access, telecom networks and services, and policies among different economies.
Hence, the cost and technology behind the sensing, device, networking and analytic solutions for the IoT should be \emph{affordable} and \emph{scale} to billions of users.
This requires reuse of commodity hardware and sensors, and existing infrastructure in novel ways rather than custom solutions with cutting-edge capabilities, or canned solutions developed for advanced economies. The cost-to-benefit trade-offs become critical.

\subsection{Whose data is it anyway ?}

The intersection of devices, communication, data and humans within IoT offers interesting incentive and business models. 
A key success of the \texttt{WWW} is the ability for businesses to monetize users' data (e.g., Google Ads using user's web data pays for free search and email services). 
With IoT, devices are going to be even more closer to humans and blend into our environment. 
Ensuring there is \emph{transparency} in data ownership, sharing, and usage is important. 
Further, there is scope for \emph{data brokering} that encourages open data sharing by users with business in return for clear rewards, be they monetary, peer recognition, or for the greater good.

\subsection{When ``good enough'' is enough ?}

IoT is naturally a diverse ecosystem with unreliability and uncertainties as: ($1$) cheap sensors mean questionable data quality, 
($2$) humans are fickle to model, ($3$) physical systems are complex, ($4$) distributed ``Things'' and intermittent communication are a given, and ($5$) data privacy puts bounds on its availability. 
As a result, analytic and decision making \emph{have} to be probabilistic; and the system and application has to be \emph{conscious} of what is ``good enough'' and not fail in the absence of perfect behavior. 

\subsection{Context determines the Action}


Given the uncertainties of the system and humans being central entities, much of the decision making within the IoT infrastructure and applications has to be contextual.
Context \emph{binds} people and ``Things'' to a common scope, and hence, will ease mining of relevant information.
There has to be semantic knowledge that captures system and social behaviour, some specified while others are learned using models. 
\emph{Intelligent agents} will often act on the behalf of humans.
They may be aware of personal preferences (e.g., Apple's Siri, Microsoft's Cortana, and Google Now), and these will interact with digital agents of service providers, utilities and vendors. 
Semantic context will have to complement web standards for structural syntax to allow such M$2$M interaction to be effective.

\subsection{Business Canvas}

If the IoT is to yield successful business models, we first need to recognize that IoT is \emph{not} a new product or market. 
What IoT brings is an \emph{additional set} of technologies, lower power, more computation and storage, cheaper devices, better wireless connectivity, much more granular control and observation capabilities. What it enables is scaling in both directions -- up and down, and the ability to look at ourselves and the world in an unprecedented degree of detail.

IoT business models fall into \emph{two} broad categories: ($1$) horizontals, concerned with enabling components and technology; and ($2$) verticals, which integrate these technologies to supply an end user with a value proposition. 

The \emph{first} set of horizontal business models is the development of specific sensors and actuators that enable the generation of new, or more cost effective observations. 
The \emph{second} model is the deployment horizontal, a business model that addresses the needs of building out to scale of the data gathering, data storage and data curation and data brokering needs of IoT based systems. 
The \emph{third} horizontal business model addresses the needs for a portfolio of analytical techniques to convert the data gathered into actionable information. 
While the first two have been the focus of IoT's precursor technologies, IoT's scale is driving active development across the board.

Verticals will pull solutions and services across these horizontals to deliver final end customer value. 
The emphasis here will be on the necessary domain and system integration expertise and the ability to build the necessary collaborations across customers and suppliers.

\section{NexGen I{\smaller o}T: Architecture} \label{sec:next-gen-architecture}

\begin{figure}[t]
\begin{center}
\includegraphics[width=3.3in]{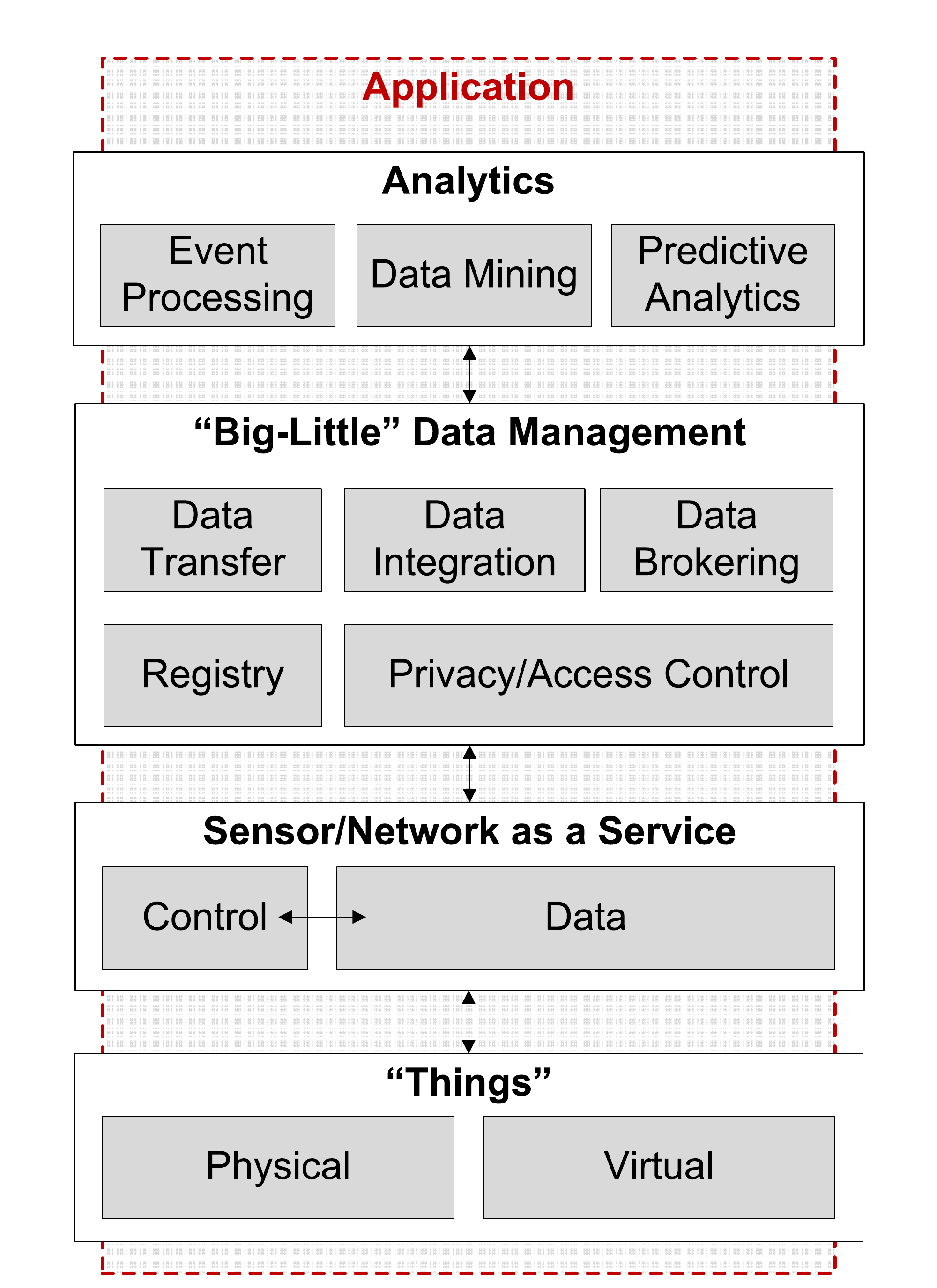}
\end{center}
\vspace{-2mm}
\caption{A High level reference architecture for the I{\smaller o}T.}
\label{fig:nexgen-iot-arch}
\end{figure}

IoT has a large application space and is driven by a wide variety of use cases.
A \emph{modular, scalable} architecture that supports adding or removing capabilities depending on the functional requirements will, therefore, be useful. 
In this section, we present one possible architecture that reflects the design principles outlined in the previous section.
Fig.~\ref{fig:nexgen-iot-arch} illustrates the $4$-layer architecture stack that we envision. 

\subsection{Physical/Virtual Space}

The lowest layer is a collection of sensing elements, or data generators and consumers that provide \emph{context} information.
This information is sensed not only from the physical space of hardware-level sensors, but also from soft sensors that exist in the virtual space.
\newline
\indent
Physical modules/platforms contain the interface to the physical world.
There exists a large number of sensor products, from multiple manufacturers, to measure various physical parameters such as temperature,
pressure, humidity, illumination, acoustics, motion, location, touch, etc.,.
These devices range in heterogeneity and complexity, from an embedded $8$-bit SoC unit with a single sensor/actuator to $32$/$64$ bit computing platform with many transducers; and interface with different proprietary communication technologies (direct Ethernet, WiFi, BLE, NFC, Zigbee, $6$LowPAN, UART or serial lines, SPI or I$2$C wired buses) over a wide variety of protocols.
\newline
\indent
The virtual sensing space is also highly diverse.
It consists of sensing entities that can be human (e.g., crowdsourced data, collaborative projects, blogs, content communities, social networking sites, virtual game worlds, virtual social worlds, electronic calendars, a travel-booking systems), their virtual agents (e.g., Siri, Cortana), or digital applications and services that aggregate and offer higher order sensing (e.g., averages from multiple physical sensors).

\subsection{Sensor/Network as a Service}


The SNaaS layer consists of control and data planes.
The data plane will offer channel models (e.g., event driven, sample and hold, etc.,) as a universal abstraction for input/output to the physical and virtual space of the sensing layer.
The control plane will be responsible for managing the sensor/ network and providing a standardized life-cycle for the discovery, configuration and use of the channel models.
The combination enables the creation of the plug-n-play infrastructure (across platforms from multiple vendors) necessary for interoperability and successful deployment of large-scale systems.
\newline
\indent
The enabling feature for plug-n-play would be a transducer electronic data sheet (TEDS) that will be used to describe the physical elements such as transducer identification, calibration, correction data, measurement range, and manufacturer related information, etc.,.
This information will either be broadcasted by the new sensing entity to the infrastructure elements, or actively pulled by the infrastructure itself from a TEDS repository on receipt of the sensor identifier. 
Providing a mechanism to describe the transducers will allow the application (user, service, agent) to record its properties and semantic description (semantic metadata). 
The control plane will expose ways to turn on/off sensors or specific attributes, change their sampling interval/transmission interval, the type of quality controls done (e.g., linear interpolation), etc.,.
\newline
\indent
Following the configuration of the ``Thing'', the data plane will publish the observations of each sensor as individual data streams to the upper layer.
The communication between the SNaas and upper layers will follow a stateless and self-describing interface using REST based protocols.
\newline
\indent
The combination of semantic capability and data push/pull (with a publish/subscribe model) will allow ``Things'' to be assembled into a semantically linked data flow graphs.
The SNaaS layer will, therefore, act as a registry for semantic discovery and linkage of ``Things'', and for virtualizing the real world.

\subsection{``Big-Little'' Data Management}

SNaaS offers a conduit for streaming ``little'' data from distributed physical and virtual
sensors. This data has to be cataloged, curated and if necessary, persisted and aggregated, in order
to perform subsequent analytics. The data management layer helps discover and maintain a \emph{Registry} of
data sources and their characteristics such as periodicity, liveliness, and quality, and make them
available for subsequent analysis. These may be from physical sensors that are deployed as part of
static infrastructure, mobile devices that may emit data transiently, or virtual sensors that share
posts based on social conditions. The registry itself may be federated using a P2P model, i.e. the
data management layer is itself distributed and not necessarily a centralized one hosted exclusively in
the Cloud.  Similarly, access to data may be as streams of events pushed using a \emph{publish-subscribe}
model, pulled on-demand, or accumulated at the device for local operations pushed to it, using existing web
and open standards like REST, COAP, Atom and JMS. These may
be controlled using \emph{data privacy} and distributed access control policies that are enforced by
this layer. 

Data management should also help integrate with aggregate and slow-changing data ``Big''
that is available from canonical sources or have accumulated from sensor streams. Often, the Big
data is hosted at central locations, globally or at local caches, to avoid costly data movement and
replication, and may have less restrictive privacy requirements. These may include
institutional
and crowd-sourced data.

Integrating this ``Big'' data with the ``Little'' data requires common vocabulary that help bridge the
semantics. DBPedia and SWEET offer reusable \emph{semantic ontologies} for general purpose and specific
domains to automate matchmaking of data requirements, while simple glossaries or taxonomies might be
adequate in other cases. Similarly, data quality checks and validation can help understand the
usability of data for specific needs.

A \emph{data brokering service} that helps interface data consumers with data producers is enabled using
these building blocks of data ownership, incentives to collect data, data description, data quality
and access control. The broker can help incentivize data collection and reuse through attribution,
barter, or even monetary rewards.

\subsection{Analytics and Decision Making}

Analytics for IoT applications often fall into categories of data and pattern mining, predictive
analytics and forecasting, event and pattern detection, and optimizations. While there may be
domain/application specific analytic, it is useful to have key analytics algorithms available that
can be reconfigured to suit common needs. Time series and regression tree forecasting are useful for
predicting future conditions, such as power demand, based on past behavior of univariate or
multivariate features. Pattern mining and clustering can help group environmental conditions and
users who exhibit similar behavior so that collective action can take place, or extrapolation from
one entity in the cluster to rest can happen, such as recommendations. Mining can also help identify
causality between a pattern of features and an event of interest. Such patterns and predictions can
help feed into realtime complex event pattern matching that detect situations of interest and help
respond to, or preferable, preempt them. These also feed into optimization algorithms that can
control physical or virtual ``Things'' to ensure reliability and efficiency of the system.

Such analytics algorithms and platforms, as before, have to operate in a distributed
environment rather than assume centralized availability of data or the ability to fit in
memory. Based on the privacy and communication constraints, and availability of distributed 
data and devices, these analytics will have to run across edge devices and the Cloud. Latency is
another key metric since analytics and optimizations over streaming data may drive realtime decision
making. Such a constrained environment in the presence of potentially unreliable data and
computation capability means that analytics and decisions will have to be \emph{probabilistic}. The
cost of performing the decision, including data, network, compute and intellectual property, need to
be factored in and traded off against the value from the decision.



\section{Discussion} \label{sec:discussion}

The high level IoT vision of building an ubiquitous society of people and ``Things'', apart from the set of enabling technologies, requires an elastic, scalable architecture.
With the continued increase in the proliferation of ``Things'', there exists a large population of physical entities that are currently observable, but do not easily connect in a meaningful way to people or to each other at the envisioned scale.
In this article, we attempted to disaggregate these core problems; and offered a trajectory with a set of design paradigms and a possible architecture for a new IoT ecosystem.

\begin{small}
\bibliographystyle{unsrt}
\bibliography{IoT-Position-references} 
\balancecolumns
\end{small}

\end{document}